\documentclass[aps,prl,twocolumn,superscriptaddress]{revtex4-2}
\usepackage{graphicx}

\begin{document}

\title{A quasi-continuous exhaust scenario for a fusion reactor:\\ the renaissance of small edge localized modes}

\author{G.F. Harrer}
\affiliation{Institute of Applied Physics, TU Wien, Fusion@\"OAW, Vienna, Austria}
\affiliation{Max Planck Institute for Plasma Physics, Garching, Germany}
\author{M. Faitsch}
\affiliation{Max Planck Institute for Plasma Physics, Garching, Germany}
\author{L. Radovanovic}
\affiliation{Institute of Applied Physics, TU Wien, Fusion@\"OAW, Vienna, Austria}
\affiliation{Max Planck Institute for Plasma Physics, Garching, Germany}
\author{E. Wolfrum}
\affiliation{Max Planck Institute for Plasma Physics, Garching, Germany}
\author{C. Albert}
\affiliation{Max Planck Institute for Plasma Physics, Garching, Germany}
\author{A. Cathey}
\affiliation{Max Planck Institute for Plasma Physics, Garching, Germany}
\author{M. Cavedon}
\affiliation{Dipartimento di Fisica "G. Occhialini'', Università di Milano-Bicocca, Milano, Italy}
\author{M. Dunne}
\affiliation{Max Planck Institute for Plasma Physics, Garching, Germany}
\author{T. Eich}
\affiliation{Max Planck Institute for Plasma Physics, Garching, Germany}
\author{R. Fischer}
\affiliation{Max Planck Institute for Plasma Physics, Garching, Germany}
\author{M. Hoelzl}
\affiliation{Max Planck Institute for Plasma Physics, Garching, Germany}
\author{B. Labit}
\affiliation{\'{E}cole Polytechnique F\'{e}d\'{e}rale de Lausanne (EPFL), Swiss Plasma Center (SPC),\\ \,\,\,\,\,CH-1015 Lausanne, Switzerland}
\author{H. Meyer}
\affiliation{CCFE, Culham Science Centre, Abingdon, Oxon, United Kingdom}
\author{F. Aumayr}
\affiliation{Institute of Applied Physics, TU Wien, Fusion@\"OAW, Vienna, Austria}
\author{the ASDEX Upgrade Team}
\affiliation{see author list of H. Meyer et al. 2019 Nuclear Fusion \textbf{59} 112014}
\author{the EUROfusion MST1 Team}
\affiliation{see author list of B. Labit et al. 2019 Nuclear Fusion \textbf{59} 086020}

\date{\today}

\begin{abstract}

Tokamak operational regimes with small edge localized modes (ELMs) could be a solution to the problem of large transient heat loads in future fusion reactors because they provide quasi-continuous exhaust while keeping a good plasma confinement. A ballooning mode mechanism near the last closed flux surface (LCFS) governed by an interplay of the pressure gradient and the magnetic shear there has been proposed for small ELMs in high density ASDEX Upgrade and TCV discharges. In this manuscript we explore different factors relevant for plasma edge stability in a wide range of edge safety factors by changing the connection length between the good and the bad curvature side. Simultaneously this influences the stabilizing effect of the local magnetic shear close to the LCFS as well as the $E \times B$ flow shear. Ideal ballooning stability calculations with the HELENA code reveal that small ELM plasmas are indeed unstable against ballooning modes very close to the LCFS but can exhibit second ballooning stability in the steep gradient region which correlates with enhanced confinement. We also present first non-linear simulations of small ELM regimes with the JOREK code including the $E \times B$ shear which indeed develop ballooning like fluctuations in the high triangularity limit. In the region where the small ELMs originate the dimensionless parameters are very similar in our investigated discharges and in a reactor, making this regime the ideal exhaust scenario for a future reactor.

\end{abstract}

\maketitle

Future devices that exploit nuclear fusion to generate electrical energy must have good particle and energy confinement to achieve enough fusion gain. In tokamaks, an excellent confinement is linked to an edge transport barrier, which is accompanied by steep pressure gradients at the edge, the so-called pedestal \cite{Wagner1982}.
However, the particle confinement must not be too good either, since in steady state operation impurities and the fusion product Helium must be removed and replaced by fresh fuel. An additional requirement for large devices is a high density at the last closed flux surface (the separatrix), a condition which must be met to protect the plasma facing components from excessive power loads \cite{Wischmeier2015}. The steep pressure gradients at the edge can be the cause of instabilities, so-called edge localized modes (ELMs), which expel both particles and energy in strong bursts. The largest and most common edge localized modes are called type-I ELMs \cite{Zohm1996a}. While still tolerable in present day machines, these type-I ELMs pose a serious threat to first wall components of future reactor-grade devices such as ITER or DEMO\cite{Riccardi2011}. The search for mitigation measures (using e.g. resonant magnetic perturbation coils or pace making of smaller ELMs by pellet injection \cite{Nazikian2015,Baylor2013}) or even better for operational regimes that avoid such strong bursts while maintaining the good confinement at high separatrix densities has been ongoing for years \cite{Viezzer2018}. 
The small ELM regime presented in this work has been observed on several machines before and was then called type-II or grassy ELM regime \cite{Ozeki1990,Saibene2005,Stober2005,Kirk2011, Oyama2006,Wolfrum2011, Xu2019}.
However, at that time the interpretation of the stability of \mbox{type-II} ELMs took place in the same parameter space as for type-I ELMs, namely edge pressure gradient and edge bootstrap current across the whole pedestal width \cite{Snyder2004}.
In the following we will demonstrate that the small ELMs, in contrast to the disastrous \mbox{type-I} ELMs, are not destabilized in the entire region of the edge transport barrier, but only in a small region just inside the last closed flux surface. Moreover, by specifically tailoring the pedestal in such a way that the small ELMs provide enough quasi-continuous transport, while reducing the pedestal width, we show that the occurrence of large type-I ELMs can be prevented without the need for any mitigation measures. In this way, we can successfully achieve a mode of operation with good confinement at high separatrix density and a wide heat load footprint in the divertor. This tokamak operating regime provides a quasi-continuous exhaust of particles and heat while preventing the occurrence of strong type-I ELMs (therefore termed quasi-continuous exhaust or QCE scenario). It shows enhanced filamentary transport \cite{Griener2020} and a significantly broadened heat-flux footprint \cite{Faitsch2020} and is therefore a particularly promising regime for future fusion devices like ITER or DEMO. 

The realization that this regime benefits from instabilities at the foot of the transport barrier \cite{Harrer2018,Labit2019}, which lead to quasi-continuous transport without significantly degrading confinement, has led us to investigate the factors that can tailor the instabilities.
These investigations show, that while the driving pressure gradient is mainly determined by density, the dominant destabilizing terms are a weak local magnetic shear, a weak poloidal flow shear and a long connection length between the bad curvature low field side (LFS) and the good curvature high field side (HFS). 
We could meanwhile experimentally demonstrate and establish the QCE-regime for a wide range of safety factors and heating powers. Moreover, ideal ballooning stability calculations show that in all discharges, the narrow region just inside the separatrix is ideally ballooning unstable. First non-linear, resistive magneto-hydrodynamic calculations of such a small ELM regime using the JOREK code \cite{Hoelzl2021} develop ballooning-like fluctuations under similar conditions such as an elevated separatrix density. We can therefore state, that the new understanding of the origin of the small ELM regime, namely a localized unstable region just inside the separatrix, is supported by several experimental findings as well as linear ideal and non-linear resistive modeling.

\begin{figure}\centering
\centering
\includegraphics[width=85mm]{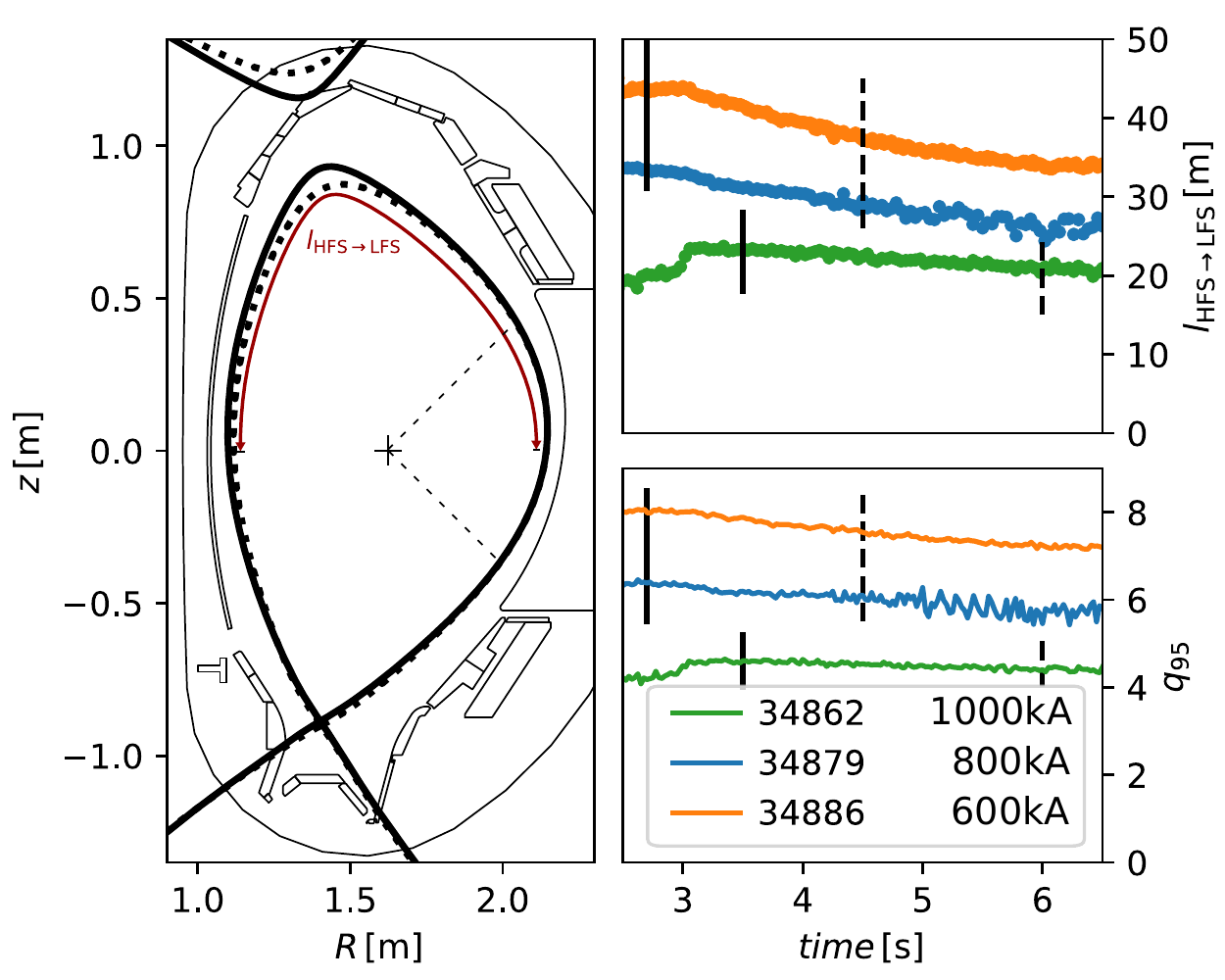}
\caption{\label{fig:shape_conlenght}Left: Cross section of the separatrix of the high (solid) and low (dashed) $l_\mathrm{HFS \rightarrow LFS}$ time points. $l_\mathrm{HFS \rightarrow LFS}$ is the length of a field line spanning from the LFS to the HFS midplane. A poloidal projection of $l_\mathrm{HFS \rightarrow LFS}$ is depicted with the red arrow. Right: temporal evolution of $l_\mathrm{HFS \rightarrow LFS}$ and the safety factor at 95\% flux throughout the three discharges.}
\end{figure}

To emphasize the broad operational range of the QCE regime this study focuses on three discharges at safety factors of $q_\mathrm{95}=4,$ 6 and 8, achieved by a plasma current variation, performed on ASDEX Upgrade close to the empirical density limit $f_\mathrm{GW,ped}>0.87$ and ITER confinement time scaling factors $H_\mathrm{98,y2}>1$. All three discharges, were programmed to have a high elongation and triangularity (solid line cross section in figure \ref{fig:shape_conlenght}, left) resulting in a small ELM dominant phase. To influence the connection length between the good and bad curvature side $l_\mathrm{HFS \rightarrow LFS}$ (measured at a poloidal radius of $\rho_\mathrm{pol}=0.99$ depicted by the red arrow in figure \ref{fig:shape_conlenght}), and therefore the ballooning stability, the plasma shapes were then altered in the same way (dotted cross section in figure \ref{fig:shape_conlenght}) for all three discharges . The temporal evolution of $l_\mathrm{HFS \rightarrow LFS}$ as well as $q_\mathrm{95}$ is depicted on the right of figure \ref{fig:shape_conlenght}.

\begin{figure}
\centering
\includegraphics[width=85mm]{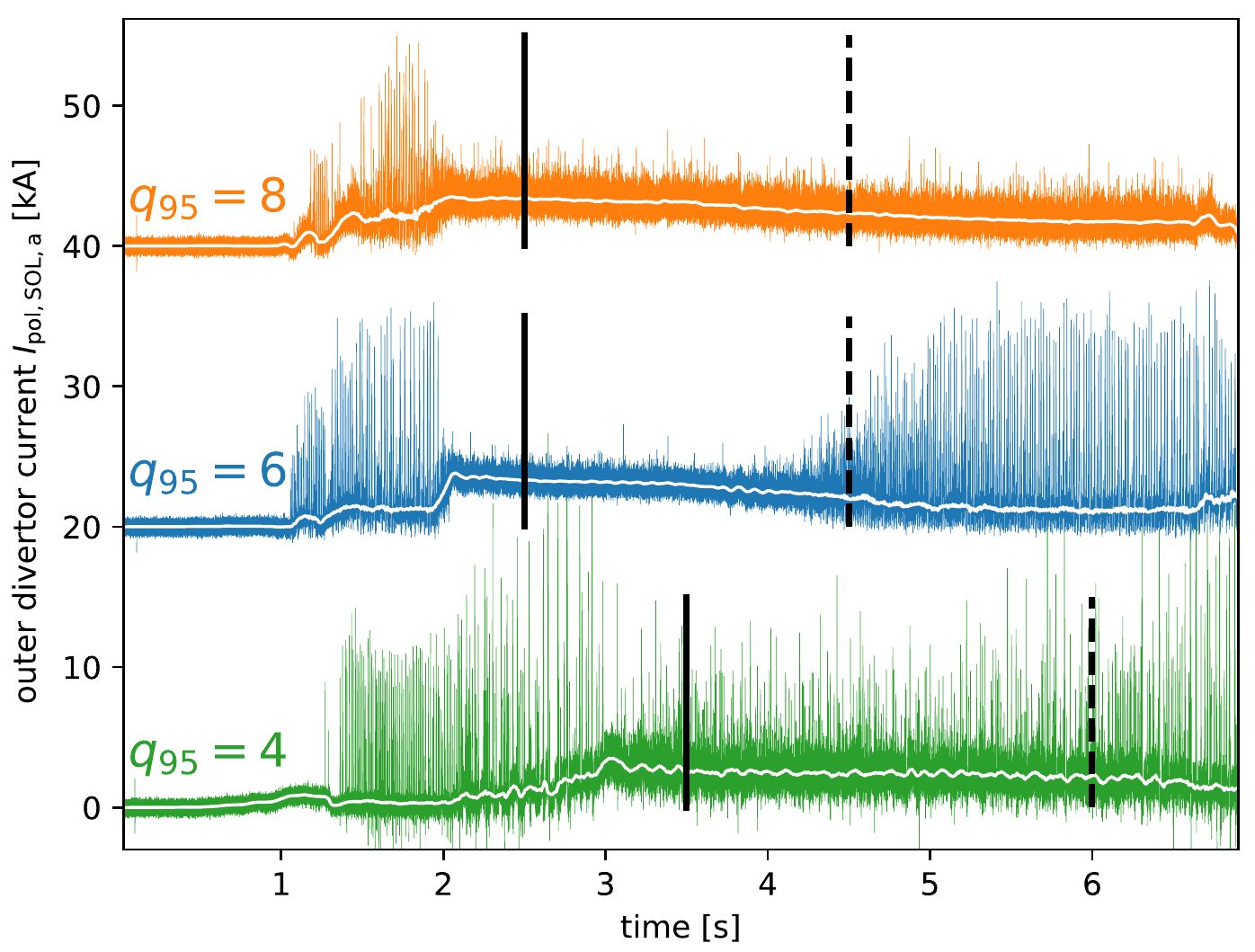}
\caption{\label{fig:ipolshift}ELM signatures of the three discharges with different edge safety factors. The two higher $q$ discharges are offset by $20\,\mathrm{kA}$ and $40\,\mathrm{kA}$ for better visibility. The solid and dashed lines mark the higher and lower connection length configurations depicted in figure \ref{fig:shape_conlenght}.}
\end{figure}

The outer poloidal divertor current $I_\mathrm{polSOLa}$, serving as an ELM indicator in metal machines, is shown in figure \ref{fig:ipolshift} for all three discharges including offsets of $20\,\mathrm{kA}$ for illustrative purposes. The small ELM regime is established at $3\,\mathrm{s}$ in the lowest $q$ discharge (\#34862) and at $2\,\mathrm{s}$ in the other two discharges. $1\,\mathrm{s}$ later, the plasma $z$ position is gradually ramped down, consequently reducing $l_\mathrm{HFS \rightarrow LFS}$. In all three discharges, at the time of the longest connection between LFS and HFS (black solid bar), the largest transport caused by the small ELMs is observable as an elevated background in the signals (white lines). With the reduction of the connection length, this background decreases in all 3 cases, whereby in the high-$q$ discharge no type-I ELMs appear, in the medium-$q$ discharge at $4.5\,\mathrm{s}$ the first type-I ELMs arise and in the low-$q$ discharge type-I ELMs occur sporadically during the entire phase, indicating that the pedestal in the low-$q$ discharge is still very close to the peeling-ballooning stability boundary.

In addition to the influence of the high-field side, there are also local mechanisms governing the small ELM stability, especially the local magnetic shear $s_l=-\mathbf{e_\perp} \cdot \mathbf{\nabla} \times \mathbf{e_\perp} $ as defined in \cite{McCarthy2013} with $\mathbf{e_\perp}=\frac{\mathbf{\nabla\Psi}}{\|\nabla\Psi\|}\times \frac{\mathbf{B}}{\| B \|} $. $s_l$ represents the local tilt of neighboring flux tubes and can stabilize ballooning modes, in particular the ones at the pedestal bottom causing the small ELMs at the LFS.

\begin{figure}\centering
\centering
\includegraphics[width=85mm]{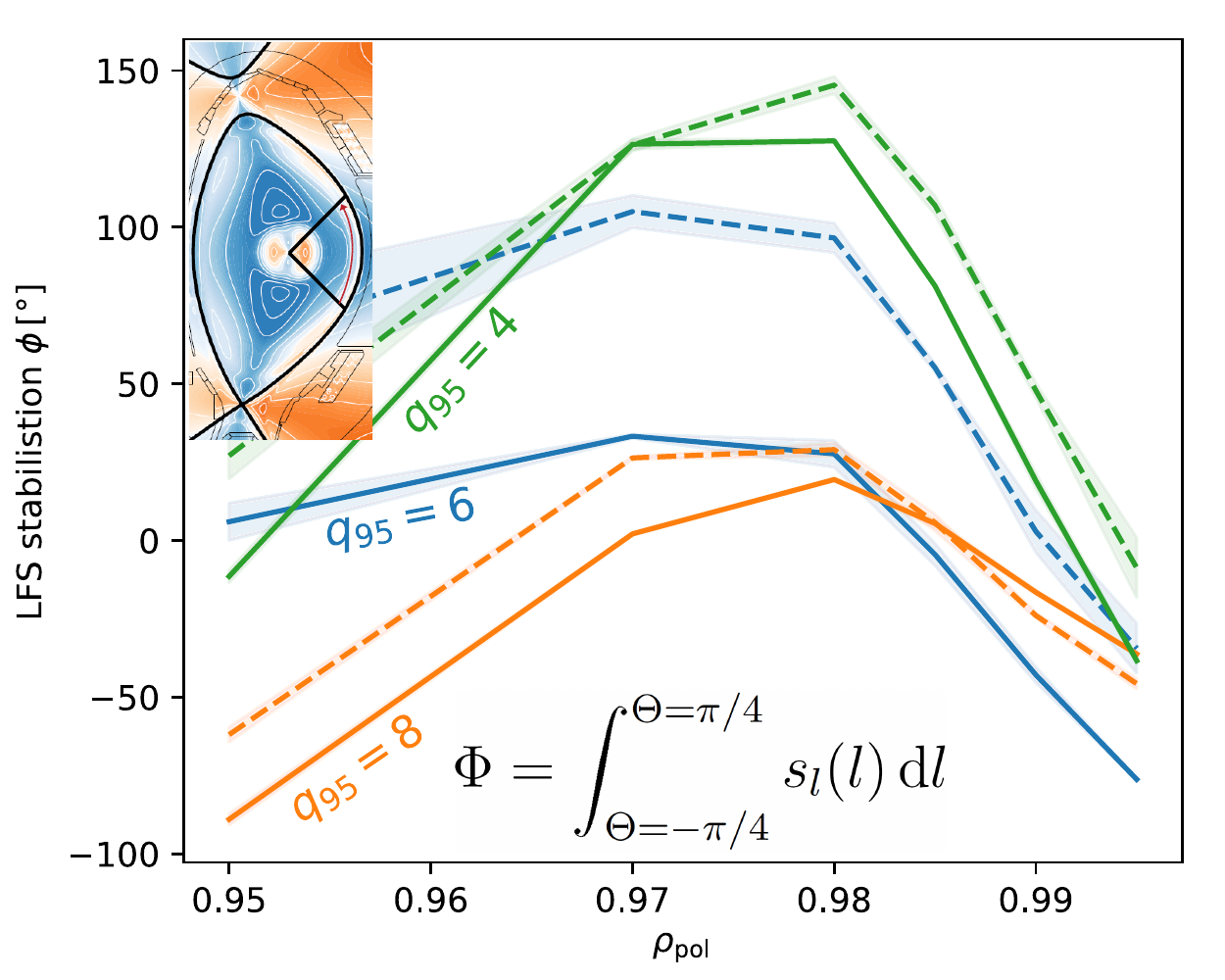}
\caption{\label{fig:slocprofs} Shear stabilization at the bad curvature side represented by the field line integrated local magnetic shear from $-45^\circ$ to $45^\circ$ w.r.t the outboard midplane. Solid lines show the high and dashed lines the lower $l_\mathrm{HFS \rightarrow LFS}$ phases. Different colors represent the different safety factors.}
\end{figure}

Figure \ref{fig:slocprofs} shows a measure of the shear stabilization of the bad curvature region around the LFS midplane achieved by integrating the local magnetic shear poloidally along the fieldlines  from $-45^\circ$ to $45^\circ$ w.r.t the midplane at different radial positions. Here the solid lines represent the high connection length phases while the lowered connection length phases are depicted with dashed lines. In the high-$q$ case, which stays in the small ELM regime, lowering the connection length only changes the stabilization inside $\rho=0.98$. The medium and high-$q$ cases show an increase of the shear stabilization at the separatrix which is in agreement with the observed transport behavior, i.e. a stabilization of the small ELMs located at the pedestal foot.

\begin{figure}\centering

\includegraphics[width=80mm]{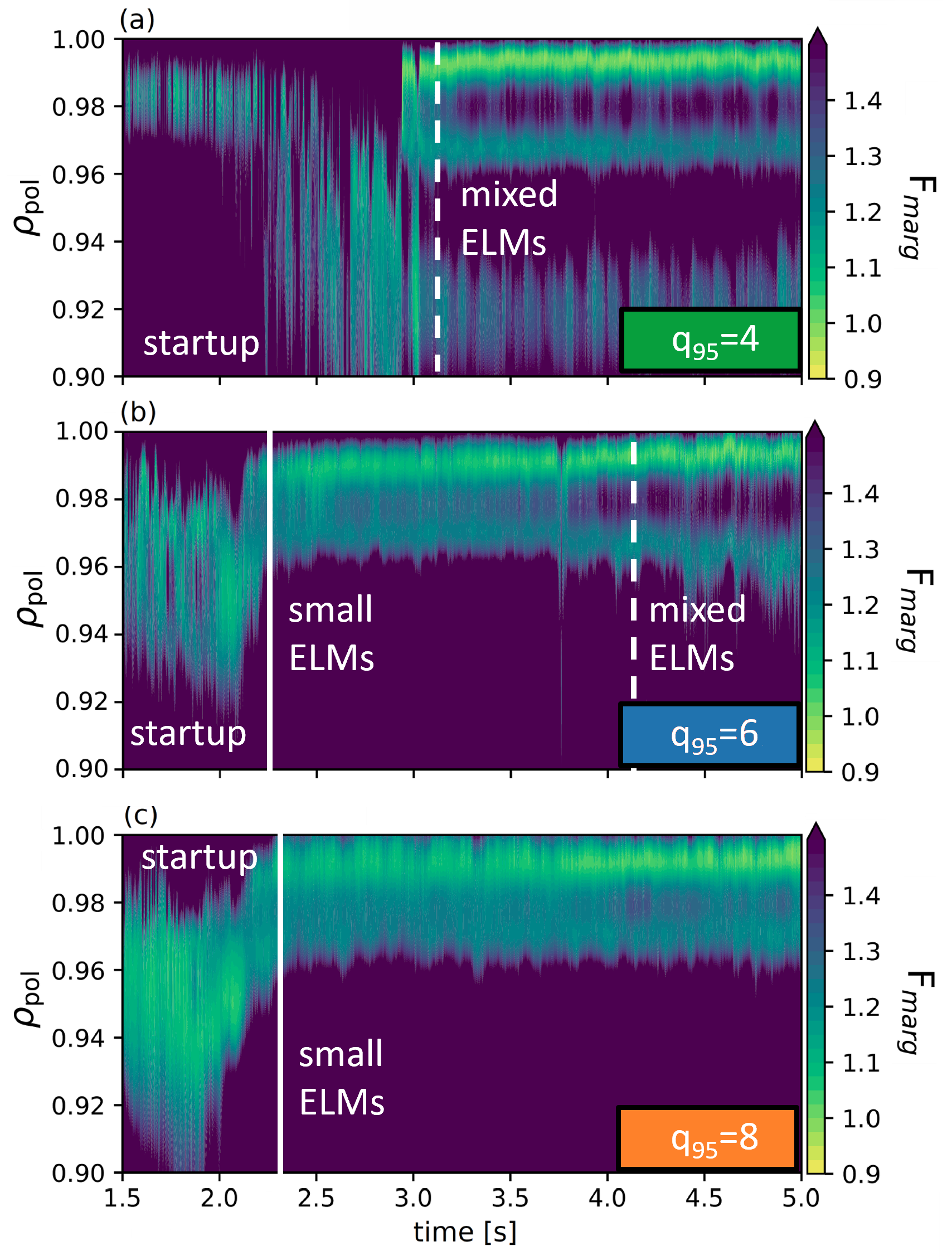}
\caption{\label{fig:helena}Time traces of the marginal stability $F_\mathrm{marg}$ of three QCE discharges for different $\rho_\mathrm{pol}$. The transitions of dominant ELM type are marked with white vertical lines. Ballooning stability is represented by dark blue colors while instability is colored in light green.}

\end{figure}

The ballooning stability of the three discharges was calculated with HELENA \cite{Huysmans1991} using high resolution IDE \cite{Fischer2016} equilibria to guarantee an accurate bootstrap current evolution. HELENA, an ideal $n\rightarrow \infty$ code calculates the linear stability at each flux surface separately which results in profiles for the critical normalized pressure gradient $\alpha_\mathrm{crit}$ at which the plasma would become ballooning unstable.
The temporal evolution of the marginal stability $F_\mathrm{marg}=\alpha_\mathrm{exp} / \alpha_\mathrm{crit}$  of the three discharges is plotted for different radial positions and shows some distinct features that can be summarized as follows: When the discharge evolves to the high shaping ($3\,\mathrm{s}$ for 34862, $2\,\mathrm{s}$ for the higher $q$ discharges), the plasma becomes ideal ballooning unstable close to the separatrix at $\rho=0.99$. In the steep gradient region at $\rho=0.98$, the plasmas are more stable (scans in $s$-$\alpha$ space show second stability access). After the connection length is lowered, the outer unstable region gets shifted even further outward and the region of stability becomes broader. The effect is best seen in the medium $q$ shot (figure \ref{fig:helena}(b)) where it leads to a broadening of the pedestal and a reappearance of type-I ELMs resulting in a mixed ELM regime. The lower LFS shear stabilization as well as the higher $l_\mathrm{HFS \rightarrow LFS}$ in the high q case keep the plasma in the small ELM regime.

The third quantity crucial for edge stability is the radial electric field and the associated flow shear, the $E \times B$ shear, or in other words the radial change of the $E \times B$ velocity.
Its main ingredient is the gradient of the radial electric field which is notoriously hard to measure experimentally.
It has been reported to suppress turbulent transport and is widely believed to be the main cause for the L-H transition \cite{Burrell1992,Wagner2007,Cavedon2020}.
To investigate its role in small ELM stability, the radial electric field $E_\mathrm{r}$ profiles for the two higher $q$ discharges have been calculated from the line intensity and the poloidal and toroidal velocity measured by the AUG CXRS diagnostics using the radial force balance. The profile reconstruction for the lowest $q$ discharge showed a minimum of $E_\mathrm{r}$ outside the separatrix, suggesting that the measurements were affected  by the type-I ELMs. The discharge has therefore been left out of this analysis.

\begin{figure}\centering
    \includegraphics[width=85mm]{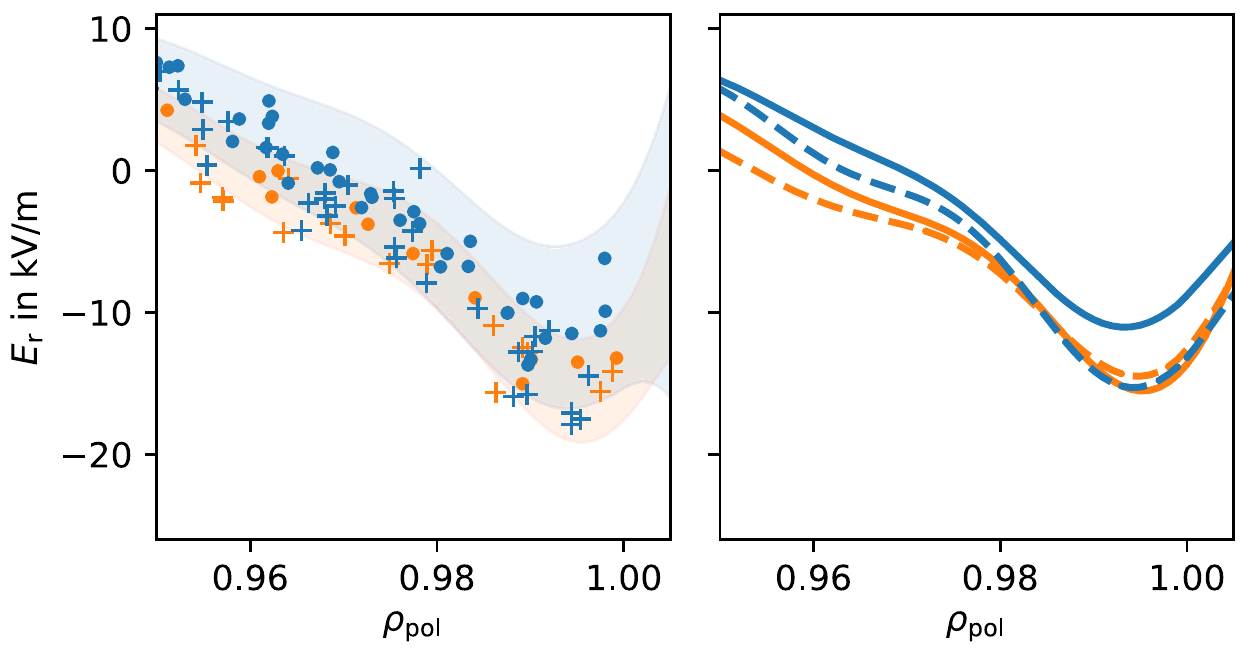}
    \caption{\label{fig:er}Radial electric field measured with CXRS of the $q_{95}=6$ (blue) and $q_{95}=8$ (orange) discharges including a Gaussian process fit of the data on the right hand side. Dots and full lines represent the high $l_\mathrm{HFS \rightarrow LFS}$ phases, while crosses and dashed lines depict lower $l_\mathrm{HFS \rightarrow LFS}$.}

\end{figure}

The data points measured are shown in figure \ref{fig:er} as circles for the higher $l_\mathrm{HFS \rightarrow LFS}$ and crosses for the lowered $l_\mathrm{HFS \rightarrow LFS}$ phases. The $E_r$ data were then fitted using the proFit Gaussian process (GP) routine \cite{profit2019} with a squared exponential kernel. The right side of figure \ref{fig:er} shows the fitted profiles for the two discharges, where again solid lines denote the higher and dashed lines the lowered $l_\mathrm{HFS \rightarrow LFS}$  phases. While radial electric field profiles in the $q_{95}=8$ case (orange) do not change with the alteration of the plasma shape, the characteristic minimum in the $E_\mathrm{r}$ profile becomes 50\% deeper in the $q_{95}=6$ case. 

The HELENA ballooning stability calculations presented in figure \ref{fig:helena} do not take the $E \times B$ shear into account. As the $E_r$ profiles and also their gradients change significantly when comparing the clean small ELM phase with the phase where type-I ELMs reappear, the influence of the $E \times B$ shear on small ELM stability cannot be disregarded. 

Recently, full type-I ELM cycles have been simulated with the non-linear resistive magneto-hydrodynamic JOREK code \cite{Cathey2020}. By using similar plasma parameters but higher separatrix density and lower heating power we were able to show simulations with small ELMs with less impact on the divertor. The simulations exhibit growth of a broad spectrum of mode numbers without a clearly dominating one and an enhanced pressure gradient close to the separatrix. With increased heating power, the $E_r$ well deepens, leading to a reoccurrence of type-I ELMs in the simulation. A detailed description of the simulation results can be found in  \cite{Cathey2021}.

The simulations point to the important role of the $E \times B$ shear, as without it, JOREK is not able to reproduce ELM cycles of a small or large kind and can only simulate single ELM events followed by unrealistically strong ballooning turbulence \cite{Orain2015}.

All three stabilizing quantities, magnetic shear, $E \times B$ shear, and the connection length between good and bad curvature regions, can and do change with the shape changes and influence the amount of transport induced by the small ELMs. The consequently changed shape of the whole pedestal appears to be the crucial factor determining the occurrence of large ELMs. It has to be stated here that the lowest $q$ case presented exhibits type-I ELMs throughout the discharge which would be disastrous for a reactor. With a higher plasma triangularity and thus an increased $l_\mathrm{HFS \rightarrow LFS}$ and lower separatrix shear stabilization, pure QCE discharges without any type-I have already been performed at AUG up to values of $q_{95}=3.6$ (using $I_\mathrm{P}=1\,\mathrm{MA}$ and $B_\mathrm{T}=2\,\mathrm{T}$). The similarity of the normalized pedestal bottom conditions in the QCE discharges to the ones that are predicted in larger devices, especially the separatrix collisionality, already motivated stability analyses of ITER equilibria which also show ballooning instability of the pedestal bottom \cite{Radovanovic2021}. Although the pressure gradient at the separatrix is hard to predict, with the separatrix densities expected to be fairly similar, the temperatures, however, higher, both at the separatrix, but especially at the pedestal top. The connection length to the stabilizing HFS is larger and could therefore play a bigger role, the flow shear, on the other hand, is proportional to $1/B$ and is therefore lower. Only a code that can take into account all these different mechanisms simultaneously will be able to make reasonable predictions.

To summarize, we have revisited a promising operational regime for magnetically confined fusion devices and analyzed the conditions under which the devices must be operated to achieve this regime, and performed (linear \& non-linear) modeling calculations to support our interpretation. The new understanding of the QCE scenario makes us confident that such a regime is the best option for future reactor-grade machines. 

\begin{acknowledgments}
G.F. Harrer is a fellow of the Friedrich Schiedel foundation for energy technology. This work has been carried out within the framework of the EUROfusion consortium and has received funding from the Euratom research and training programme 2014-2018 and 2019-2020 under grant agreement No 633053. The views and opinions expressed herein do not necessarily reflect those of the European Commission.
\end{acknowledgments}

\bibliography{harrer_prl_bib}

\end{document}